\documentstyle[preprint,tighten,aps,epsf]{revtex}
\begin{document}
\input{psfig}
\draft
\preprint{MKPH-T-96-10}
\title{
Virtual Compton Scattering off
the Nucleon in Chiral Perturbation Theory
}
\author{Thomas R. Hemmert and Barry R. Holstein}
\address{Department of Physics and Astronomy,
University of Massachusetts,
Amherst, MA  01003}

\author{Germar Kn\"{o}chlein and
  Stefan Scherer}
\address{Institut f\"{u}r Kernphysik,
Johannes Gutenberg-Universit\"{a}t,
D-55099 Mainz, Germany
}
\date{\today}
\maketitle
\begin{abstract}
We investigate the spin-independent part of the virtual Compton
scattering (VCS) amplitude off the nucleon within the framework of chiral
perturbation theory. We perform a consistent calculation to third
order in external momenta according to Weinberg's power counting. With
this calculation we can
determine the second- and fourth-order
structure-dependent coefficients of the general
low-energy expansion of the spin-averaged
VCS amplitude based on gauge invariance, crossing symmetry and the
discrete symmetries. We discuss the kinematical regime to which our
calculation can be applied and compare our expansion with the
multipole expansion by Guichon, Liu and Thomas. We
establish
the connection of our calculation with the generalized
polarizabilities of the nucleon where it is possible.
\end{abstract}
\newpage
\narrowtext
\section{Introduction}
Recently, there has been greatly increased activity in the field of virtual
Compton scattering (VCS) off the nucleon both
on the experimental \cite{ELFE,a1proposal,CEBAF,CEBAF2,Bates}
as well as on the theoretical \cite{GLT,Van,KKS1,GLT2,MD,FS1}
side. At the c.w.\ electron accelerator MAMI (Mainz) the A1
collaboration has taken initial data \cite{a1proposal},
and
proposals for similar experiments have been developed
at CEBAF (Newport News) \cite{CEBAF,CEBAF2}
or are being prepared at MIT-Bates \cite{Bates}.
Compared with ordinary (real)
Compton scattering, VCS off the nucleon offers a much greater variety of
experimental possibilities,
because the response of the hadronic system to the
electromagnetic probe can be investigated by independently varying the
energy and the momentum of the initial state photon and thus allows
for probing nucleon structure in both transverse and longitudinal
modes over a wide kinematic range, covering both the perturbative and
nonperturbative regimes.  In the former \cite{ELFE},
one can describe the VCS
process in terms of a perturbation series using the interaction vertices of
QCD with quarks and gluons as explicit degrees of freedom.  Studies
of this perturbative regime of QCD require high
resolution. Using an electromagnetic probe, this means that one has
to perform an experiment wherein the momentum transfer to the hadronic
system mediated by a virtual photon is large, so that we are in
the kinematic range of deep inelastic electron scattering.
On the other hand, a long history of experimental facts suggests
that the effective degrees of freedom of a hadronic system
at low energies are not quarks and gluons but baryons and
mesons or -
in our special case - nucleons and pions. If one performs an
electron scattering experiment which does not involve large momentum
transfer to the hadronic system, one probes the confined phase of strongly
interacting matter and does not resolve its underlying quark
structure---our paper will deal with this nonperturbative phase of a
hadronic system.  (Another interesting line of work is
probe the regime {\it between} the perturbative and
nonperturbative phases.  It
is not clear at which momentum transfer such a ``phase transition'' will
take place, and this will no doubt be addressed by future experiments. )

Despite much effort, even in the nonperturbative regime
there remain many questions
to be investigated in order to complete our picture of the nucleon
and other hadrons.  Important quantities which probe the
compositeness of a system are
its electromagnetic polarizabilities,
which characterize the response of the system to
an external electric or magnetic field \cite{88}.
In real Compton scattering, the electric and magnetic polarizabilities
of the nucleon --- $\alpha_0$ and $\beta_0$ ---
appear as the first model-dependent coefficients beyond
the low-energy theorem (LET) of Low \cite{Low} and Gell-Mann and
Goldberger \cite{GG}, and have been measured both for the
proton and the neutron \cite{77}. These polarizabilities have also
been calculated
within various models (for an overview see,
{\it{e.g.}}, \cite{Holsteinp,Lvovp}).
The LET of real Compton scattering has recently been extended to
include virtual photons in \cite{GLT,KKS1}, and the generalizations of
the polarizabilities were defined in \cite{GLT},
where also a first prediction for these ``generalized polarizabilities''
was obtained within the framework of a non-relativistic quark model.
These results were refined in \cite{GLT2} to include recoil
corrections.
In \cite{Van} an effective Lagrangian approach including
nucleon and two-pion resonance mechanisms was used to determine
the generalized electric and magnetic polarizability as a function
of the initial photon momentum --- $\alpha\left( \mid \vec q \mid
\right)$
and $\beta \left( \mid \vec q \mid \right)$.
A very recent field theoretical
calculation \cite{MD} investigates the spin-independent
polarizabilities within the framework of the linear sigma model. We
will herein utilize the technique of heavy baryon chiral perturbation
theory and compare our results with these earlier calculations.

It is well-known that the nonperturbative region can in general
be successfully described in terms of the
interaction of baryon and pseudoscalar meson degrees of freedom.
In the last decade a new approach, chiral
perturbation theory \cite{Weinberg,GL},
was developed which describes
the low-energy regime
of QCD in terms of these effective degrees of freedom (pions and
nucleons) while simultaneously requiring
the symmetries of the underlying gauge
theory and has yielded remarkable results.
It was first applied to the sector of
pseudoscalar mesons (see, {\it e.g.}, \cite{DGH} for a pedagogical
introduction) and then extended to case of pion-nucleon interactions
\cite{GSS}. The most recent version, which
is known as the ``heavy baryon formulation'' of chiral perturbation
theory \cite{JM,BKKM1,Ecker},
uses techniques, which are well-known from heavy quark
calculations.  It allows a consistent power counting scheme to be
developed, which was not possible with the former relativistic
formulation.
Real Compton scattering and the
polarizabilities of the nucleon were among the first calculations to be
performed in the nucleon sector of
chiral perturbation theory \cite{BKKM1,BKM4,BKMS,BKM1},
and the experimental
verification of those predictions provides an important test of its
validity.

In this paper we extend these
predictions to the VCS case and point out the strengths as well as
the limitations of our
approach.  After a discussion in section II
of the kinematics and the amplitude
structure for VCS, we will in section III briefly touch upon
the basic ingredients of
heavy baryon chiral perturbation theory and then proceed to calculate
the tree and loop diagrams to third order in the chiral expansion with
respect to external momenta ---
${O}(p^3)$.  In section IV we analyze the low-energy
structure of the VCS amplitude, which follows from general principles
like gauge invariance, crossing symmetry and the discrete symmetries
\cite{KKS1,FS1},
and predict the
structure-dependent constants for the {\it
spin-independent}\footnote{The analogous spin-dependent calculation
will be described in a subsequent publication.} part of the VCS
amplitude, which are beyond the predictive power of
such a general approach.  Finally, in a concluding section V we discuss the connection of
those structure constants to the
generalized polarizabilities defined in \cite{GLT} and relate the two
approaches.

\section{Kinematics and Chiral Expansion}
We begin our discussion of VCS off the nucleon by specifying
our notation for the process
\begin{equation}
\gamma^*(\varepsilon^{\mu},q^{\mu}) + N(p_i^{\mu}) \rightarrow
\gamma(\varepsilon'^{*\mu},q'^{\mu}) +
N(p_f^{\mu})\, .
\label{process}
\end{equation}
Here the nucleon four-momenta in the initial and final state are denoted
by $p_i^{\mu} = (E_i,\vec p_i)$ and $p_f^{\mu}
= (E_f,\vec p_f)$, respectively.
Since we will not in this paper discuss the component of
the VCS amplitude which depends on nucleon spin, we have omitted
indices for the nucleon spin states.  The initial (final) state photon is
characterized by its four-momentum $q^{\mu}=(\omega,\vec q\,)$
($q'^{\mu}=(\omega',\vec q{\,'})$) and polarization vector
$\varepsilon^{\mu}= ( \varepsilon^0, \vec \varepsilon\, )$
($\varepsilon'^{\mu} = ( \varepsilon'^{0}, \vec \varepsilon\,')$).
Whereas the final state photon is assumed to be real,
\begin{equation}
q'^2 = \omega'^2 - \vec q{\,'}^2 = 0 \, ,
\end{equation}
the initial state photon is taken to be space-like, {\it i.e.}
\begin{equation}
q^2 = - Q^2 < 0 \, .
\end{equation}
Since our discussion refers to an electron scattering
experiment, wherein the virtual photon is the exchanged particle between
the electron and the hadronic current, we can write the polarization
vector of the virtual photon as
\begin{equation}
\varepsilon_{\mu} = e {\bar{u}}_{e'} \gamma_{\mu} {u}_e / q^2 \, ,
\end{equation}
where $u_e$ and ${\bar{u}}_{e'}$ are the initial and final state
electron Dirac spinors,
$\gamma_{\mu}$ is a Dirac matrix (see, {\it e.g.}, \cite{BD}) and
$e = \mid e \mid \approx \sqrt{4 \pi / 137} > 0 $.
Finally, we define the Lorentz invariant momentum transfer
\begin{equation}
t = (q - q')^2 \, ,
\end{equation}
which will be useful in the following discussion.
We evaluate the VCS amplitude in the c.m. system,
\begin{eqnarray}
\vec p_i & = & - \vec q \, ,\label{pi}\nonumber\\
\vec p_f & = & - \vec q{\,'} \label{pf}\, ,
\end{eqnarray}
and for our calculation we select a special frame of reference,
\begin{eqnarray}
{\hat{q}} & = & (0,0,1) \, ,\nonumber\\
{\hat{q}}' & = & (\sin \theta,0,\cos \theta) \, ,
\end{eqnarray}
wherein we define an orthonormal set of basis vectors:
\begin{mathletters}
\begin{eqnarray}
{\hat{e}}_z & = & {\hat{q}} \, , \label{ez}\\
{\hat{e}}_y & = & {\hat{q}} \times {\hat{q}}'/\sin \theta \, ,\\
{\hat{e}}_x & = & {\hat{e}}_y \times {\hat{e}}_z \, .
\end{eqnarray}
\end{mathletters}
The complete VCS amplitude can in general be expressed in terms of three
independent kinematical quantities, {\it e.g.}, $\omega'$, $\mid
\vec q \mid$ and $\theta$ \cite{KKS1}. Note that electron scattering
kinematics implies $\omega' < \mid \vec q \mid $ which follows from
energy conservation and $Q^2>0$.  
It turns out that our choice of variables facilitates
the chiral expansion of the amplitudes, allowing chiral
power counting \cite{Weinberg} to be applied straightforwardly.  The
point here is that the
chiral expansion is an expansion in terms of external momenta $p$, where
in VCS on the nucleon the scale is set by the nucleon mass $M$ and $4
\pi F$. The pion decay constant has the value $F = 92.4 \,
{\mathrm{MeV}}$.  Now it
is a crucial point to define what is meant by the term ``external
momentum''. In the following analysis we will take the view that
{\it{each component}} of
the four-momenta of the photons and of the three-momenta of the
nucleons (see Eq.\ (\ref{pi}))
has to be ``small''
compared to $M$.\footnote{As a practical matter, since delta degrees of
freedom are not included directly, the region of applicability is more
appropriately $\omega',|\vec{q}|< m_\pi$.} As a consequence we will
count the terms $\omega'/M$
and $\mid \vec q\mid/M$ to be of the same chiral order, whereas a term
like $\mid \vec q \mid^2/M^2$ is suppressed by one chiral order with
respect to the two preceding terms. This has important consequences: Let
us, {\it e.g.}, consider the difference of the initial and final state photon
energies, which occurs frequently in the subsequent
heavy baryon calculation. From energy conservation one obtains
\begin{equation}
\omega - \omega' = \frac{\omega'^2}{2M} - \frac{\mid \vec q
  \mid^2}{2M} + O\left(\frac{r^4}{M^3}\right) \, , \label{kinex}
\end{equation}
where $r$ stands for either $\omega'$ or $\mid \vec q \mid$. From
Eq.\
(\ref{kinex}) we infer that we must count the difference of the initial and
final state photon energies as being one order higher than both energies
themselves.
Now imagine that we have evaluated a diagram
which is of the chiral order $O \left(p^n \right)$
and that part of this amplitude is
proportional to $\omega - \omega'$.  We find that this component of the
amplitude can then be neglected because it contributes
at the same order as a genuine
$O \left( p^{n+1} \right)$ diagram.
Specifically we will find below
that the loop diagrams we wish to calculate are already
$O \left( p^3 \right)$ and
contain many pieces which are proportional to $\omega - \omega'$.
Such contributions vanish from the beginning
in the case of real Compton scattering, and we can
also neglect such terms in our ${O}(p^3)$ VCS calculation,
because they would contribute at the same order as a $O \left(p^4 \right)$
contribution, which is not addressed in this paper.  Finally, it is
useful to present the expansion of $q^2$ and $t$ in terms of the
three independent kinematical quantities,
\begin{mathletters}
\begin{eqnarray}
q^2 & = & \omega'^2 - \mid \vec q \mid^2 +
{{O}}\left(\frac{r^3}{M}\right) \, ,
\label{q2}\\
t & = & - \omega'^2 - \mid \vec q \mid^2 + 2 \omega' \mid \vec q \mid
\cos \theta + {{O}}\left(\frac{r^3}{M}\right) \, ,
\label{t}
\end{eqnarray}
\end{mathletters}
which we will also use below.

\section{Calculation in Heavy Baryon Chiral Perturbation Theory}

In this section we will extend the $O \left( p^3 \right)$ chiral heavy baryon
calculation for real Compton scattering \cite{BKKM1,BKM1}
to the case where $q^2
< 0$. In the case of real Compton scattering in the forward direction,
a $O \left( p^4 \right)$
calculation has been
performed in \cite{BKM4,BKMS},
which yields only small corrections to the $O \left( p^3 \right)$
result. Guided by this observation, we expect our $O \left( p^3 \right)$
calculation to
provide a reasonable estimate in the kinematical region we are considering.
The invariant amplitude for VCS can be written in the form \cite{BD}
\begin{equation}
{\cal{M}}_{VCS} = - i e^2 \varepsilon_{\mu} M^{\mu} \, = - i e^2
\varepsilon_{\mu} \varepsilon'^*_{\nu} M^{\mu \nu} \, . \label{invamp}
\end{equation}
We will utilize the Coulomb gauge,
\begin{equation}
\varepsilon'^{\mu} = \left( 0, \vec
  \varepsilon\,' \right)\, , \,\,\,\,\, \vec \varepsilon\,'
\cdot \vec q\,' = 0
\end{equation}
for the
real photon.
We decompose the space components of the virtual photon
polarization vector $\varepsilon^{\mu}$
into a purely transverse
and a purely longitudinal part in our frame of reference,
\begin{eqnarray}
\vec \varepsilon & = & \vec \varepsilon_T + \vec \varepsilon_L\, , \\
\vec \varepsilon_T & = & \varepsilon_x {\hat{e}}_x + \varepsilon_y
{\hat{e}}_y\, ,\\
\vec \varepsilon_L & = & \varepsilon_z {\hat{e}}_z \, . \label{epsl}
\end{eqnarray}
Technically we calculate ${\cal{M}}_{VCS}$ with an initial photon
polarization vector
\begin{equation}
a^{\mu} = \left( 0, \vec a \right)\, , \,\,\,\,\,
\vec a = \vec \varepsilon_T + \frac{q^2}{\omega^2} \vec \varepsilon
\cdot {\hat{q}} {\hat{q}} =
\vec \varepsilon_T + \frac{q^2}{\omega^2} \varepsilon_z {\hat{q}}
\label{a}
\end{equation}
with $q \cdot a \neq 0$ \cite{AFF},
which simplifies
the calculation by substantially reducing the number of Feynman
diagrams which must be considered.
The invariant amplitude can also be decomposed
into a
transverse and a longitudinal part.
Using current conservation,
\begin{equation}
q_{\mu} M^{\mu} = 0 \, ,
\end{equation}
Eq.\ (\ref{invamp}) can be written as
\begin{equation}
{\cal{M}}_{VCS} = i e^2 \left( \vec \varepsilon_T \cdot \vec M_T +
  \frac{q^2}{\omega^2} \varepsilon_z M_z \right) \, .
\end{equation}
The transverse part of the invariant amplitude consists of
eight independent structures.
We will use a notation similar to that given in \cite{BKM1},
\begin{eqnarray}
\vec \varepsilon_T \cdot \vec M_T & = & {\vec{\varepsilon}}\,'^* \cdot
{\vec{\varepsilon}}_T
A_1
+ {\vec{\varepsilon}}\,'^* \cdot {\hat{q}} {\vec{\varepsilon}}_T
\cdot {\hat{q}}'
A_2 \nonumber\\
&+& i \vec \sigma \cdot \left( {\vec{\varepsilon}}\,'^*
\times {\vec{\varepsilon}}_T
 \right) A_3
+ i \vec \sigma \cdot \left( {\hat{q}}' \times {\hat{q}} \right)
{\vec{\varepsilon}}\,'^* \cdot {\vec{\varepsilon}}_T A_4 \nonumber \\
&+& i \vec \sigma \cdot \left( {\vec{\varepsilon}}\,'^*
\times {\hat{q}} \right)
{\vec{\varepsilon}}_T \cdot {\hat{q}}' A_5
+ i \vec \sigma \cdot \left(
{\vec{\varepsilon}}\,'^* \times {\hat{q}}' \right)
{\vec{\varepsilon}}_T \cdot {\hat{q}}' A_6 \nonumber \\
&-& i \vec \sigma \cdot \left( {\vec{\varepsilon}}_T \times {\hat{q}}' \right)
{\vec{\varepsilon}}\,'^* \cdot {\hat{q}} A_7
- i \vec \sigma \cdot \left( {\vec{\varepsilon}}_T \times {\hat{q}} \right)
{\vec{\varepsilon}}\,'^* \cdot {\hat{q}} A_8 \, , \label{trans}
\end{eqnarray}
where $\sigma_i \,\,\,\, \left(i \in \left\{ x,y,z \right\} \right)$
are the Pauli spin matrices.
(Note that in the special case of real Compton scattering time reversal
invariance
imposes two additional
constraints on the amplitudes, as a result of which we find
$A_5 = A_7$ and $A_6 =
A_8\,$.)  Only two of these amplitudes --- $A_1$ and $A_2$ --- are
independent of nucleon spin.
For the longitudinal component one finds four independent structures,
\begin{eqnarray}
M_z & = & {\vec{\varepsilon}}\,'^* \cdot {\hat{q}} A_9
+ i \vec \sigma \cdot \left( {\hat{q}}' \times {\hat{q}} \right)
{\vec{\varepsilon}}\,'^* \cdot {\hat{q}} A_{10}
\nonumber \\
& & + i \vec \sigma \cdot {\vec{\varepsilon}}\,'^* \times {\hat{q}} A_{11}
+ i \vec \sigma \cdot {\vec{\varepsilon}}\,'^* \times {\hat{q}}' A_{12} \,
, \label{long}
\end{eqnarray}
and in this case a single amplitude --- $A_9$ --- is spin-independent.
In this paper, we restrict our consideration
to this spin-independent component of the matrix element,
\begin{eqnarray}
&&{\cal{M}}_{VCS}^{non-spin} = {\cal{M}}_{VCS} -
{\cal{M}}_{VCS}^{spin}
\nonumber\\
&=&
i e^2 \left[
\vec \varepsilon\,'^*
\cdot \vec \varepsilon_T A_1 + \vec \varepsilon\,'^* \cdot {\hat{q}}
\vec \varepsilon_T \cdot {\hat{q}}' A_2 + \vec \varepsilon\,'^*
\cdot {\hat{q}}
\frac{q^2}{\omega^2} \varepsilon_z
A_9 \right] \, ,
\end{eqnarray}
which can also be obtained from the full amplitude ${\cal{M}}_{VCS}$
as
\begin{equation}
{\cal{M}}_{VCS}^{non-spin} = \frac{1}{2} {\mathrm{tr}} \left(
  {\cal{M}}_{VCS} \right).
\end{equation}

We calculate
${\cal{M}}_{VCS}^{non-spin}$ using the standard chiral perturbation theory
Lagrangian in the heavy baryon formulation to $O \left( p^3 \right)$ in
the nucleon sector \cite{BKKM1,Ecker},
\begin{equation}
{\cal{L}}_{\pi N} = {\cal{L}}_{\pi N}^{(1)} + {\cal{L}}_{\pi N}^{(2)}
+ {\cal{L}}_{\pi N}^{(3)} \, ,
\end{equation}
with
\begin{mathletters}
\begin{eqnarray}
{\cal{L}}_{\pi N}^{(1)} & = & \bar N_v(iv \cdot D + g_A S \cdot
u) N_v \, , \label{L1}\\
{\cal{L}}_{\pi N}^{(2)} & = &  - \frac{1}{2M} \bar N_v \left\{
  \vphantom{\frac{1}{2}} D \cdot
D -(v\cdot D)^2 \right. \nonumber \\
&-&
\left.\frac{1}{2} \varepsilon_{\mu \nu \rho \sigma} v^{\rho} S^{\sigma}
\left[ f_+^{\mu \nu} \left( 1 + 4 c_6 \right)
+ 2 v^{(s),\mu \nu} \left( 1 + 2 c_7 \right) \right] \right\} N_v
\, , \label{L2}\\
{\cal{L}}_{\pi N}^{(3)} & = & \frac{1}{2 M^2} \bar N_v
\left\{ \left[ f_+^{\mu \nu} \left(c_6 + \frac{1}{8} \right)
+ v^{(s),\mu \nu} \left( c_7 + \frac{1}{4} \right) \right]\right.\nonumber\\
&\times&\left.\varepsilon_{\mu\nu\rho\sigma} 
S^{\sigma} i D^{\rho} + \mathrm{h.c.}
\vphantom{\left(\frac{1}{8}\right)}
\right\} N_v
\, ,
\label{piN}
\end{eqnarray}
\end{mathletters}
where $\varepsilon_{0123} = 1 $.
We have only kept those
terms which contribute to a $O \left( p^3 \right)$
VCS calculation. In particular terms linear in the photon fields,
which vanish in our gauge, have been omitted.  Moreover, we note that
${\cal{L}}^{(3)}_{\pi N}$ contributes only to the spin-dependent piece
of the VCS amplitude and is irrelevant for the spin-independent part.
The velocity-dependent field $N_v$ is projected out from
the nucleon Dirac spinor $\Psi_N$,
\begin{equation}
N_v = {\rm{exp}} \left[ i M v \cdot
  x \right] P_v^+ \Psi_N \, ,
\end{equation}
where the projection operator is given by
\begin{equation}
P_v^+ = \frac{1}{2} \left( 1 + \not\!{v} \right) \, .
\end{equation}
where $v^\mu$ is a velocity vector satisfying $v^2 = 1$.
The covariant derivative is defined as
\begin{equation}
D_{\mu} = \partial_{\mu} + \Gamma_{\mu} - i v_{\mu}^{(s)} \, ,
\end{equation}
where
\begin{mathletters}
\begin{eqnarray}
\Gamma_{\mu} & = & \frac{1}{2} \left\{ u^{\dagger} \left(
    \partial_{\mu} - i r_{\mu} \right) u + u \left( \partial_{\mu} - i
    l_{\mu} \right) u^{\dagger} \right\} \, , \\
u & = & U^{\frac{1}{2}} \, , \\
U & = & {\mathrm{exp}} \left(i \vec \tau \cdot \vec \pi /
F \right)\, . \label{eq:aa}
\end{eqnarray}
\end{mathletters}
In Eq. (\ref{eq:aa}) $\vec \tau$ are the conventional Pauli isospin
matrices, while $\vec \pi$ represents the interpolating pion field.
The field strength tensors are defined as
\begin{mathletters}
\begin{eqnarray}
v_{\mu \nu}^{(s)} & = & \partial_{\mu} v_{\nu}^{(s)} - \partial_{\nu}
v_{\mu}^{(s)} \, , \\
f_+^{\mu \nu} & = & u F_L^{\mu \nu} u^{\dagger} + u^{\dagger} F_R^{\mu
  \nu} u \, , \\
F_R^{\mu \nu} & = & \partial^{\mu} r^{\nu} - \partial^{\nu} r^{\mu} -
i \left[ r^{\mu}, r^{\nu} \right] \, , \\
F_L^{\mu \nu} & = & \partial^{\mu} l^{\nu} - \partial^{\nu} l^{\mu} -
i \left[ l^{\mu}, l^{\nu} \right] \, ,
\end{eqnarray}
\end{mathletters}
and to the order considered the corresponding coefficients are related
to the anomalous magnetic moments,
\begin{mathletters}
\begin{eqnarray}
c_6 & = & \frac{\mu_V}{4} - \frac{1}{4} = \frac{1}{4} \left( \mu_p -
  \mu_n - 1 \right) \, , \\
c_7 & = & \frac{\mu_S}{2} - \frac{1}{2} = \frac{1}{2} \left( \mu_p +
  \mu_n - 1 \right) \, .
\end{eqnarray}
\end{mathletters}
In our application the right- and left-handed currents, $r_{\mu}$ and
$l_{\mu}$,  are the
isovector part of the electromagnetic current and $v_{\mu}^{(s)}$ is
the isoscalar piece,
\begin{eqnarray}
r_{\mu} & = & - \frac{e}{2} A_{\mu} \tau_3 \, , \nonumber\\
l_{\mu} & = & - \frac{e}{2} A_{\mu} \tau_3 \, , \nonumber\\
v_{\mu}^{(s)} & = & - \frac{e}{2} A_{\mu} \, .
\end{eqnarray}
The Pauli-Lubanski spin vector is given as
\begin{equation}
S^{\mu} = \frac{i}{2} \gamma_5 \sigma^{\mu \nu} v_{\nu}
\end{equation}
with $\sigma^{\mu \nu} = \frac{i}{2} \left[ \gamma^{\mu}, \gamma^{\nu}
  \right]$.
Finally, we define the quantity
\begin{eqnarray}
u_{\mu} & = & i \left\{ u^{\dagger} \left( \partial_{\mu} - i r_{\mu}
    \right) u - u \left( \partial_{\mu} - i l_{\mu} \right)
    u^{\dagger} \right\}\,.
\end{eqnarray}
In the heavy baryon calculation we can write the initial and final
nucleon
momenta as
\begin{eqnarray}
p_i^{\mu} & = & M v^{\mu} + t_i^{\mu} \, , \nonumber \\
p_f^{\mu} & = & M v^{\mu} + t_f^{\mu} \, ,
\end{eqnarray}
where $t_i^{\mu}$ and $t_f^{\mu}$ denote off-shell
momenta.  We choose $v^{\mu} = (1,0,0,0)$ such that $v \cdot a = 0$, where the
polarization vector $a^{\mu}$ is given in Eq. (\ref{a}).

In the pion sector we can restrict ourselves to the $O \left(p^2
\right)$
Lagrangian,
\begin{equation}
{\cal{L}}_{\pi \pi}^{(2)} = \frac{F^2}{4} {\mathrm{tr}}
\left[ \left(\nabla_{\mu} U\right)^{\dagger} \nabla^{\mu}
U \right] \, \label{Lpi},
\end{equation}
where
\begin{equation}
\nabla_{\mu} U = \partial_{\mu} U + \frac{1}{2} i e A_{\mu} \left[
  \tau_3, U \right].
\end{equation}
represents the covariant derivative.  Here
we have omitted the usual mass term, because we only need to consider
pion-photon vertices generated by ${\cal{L}}_{\pi \pi}^{(2)}$.  We
also observe that
power counting arguments, in principle,
require the inclusion of a tree diagram
with a $\pi^0$ in the $t$-channel in our calculation
(Fig. \ref{treefig} (d)). For this diagram
we need the lowest order Wess-Zumino Lagrangian \cite{WZ,Witten},
which has odd intrinsic parity and is $O \left( p^4 \right)$.
However, this
diagram does not contribute to the spin-independent part of the VCS
amplitude, ${\cal{M}}_{VCS}^{non-spin}$, but only to its
spin-dependent
counterpart,
${\cal{M}}_{VCS}^{spin}$. For this reason we will not discuss it in the
following.
Finally, we wish to emphasize that we do not require any
additional diagrams compared to the calculation for real Compton
scattering \cite{BKKM1}. The
complete set of diagrams we have to
calculate is given in Figs. \ref{treefig} ((a) $s$-channel, (b)
$u$-channel,
(c) contact diagram)
and \ref{loopfig} (loop diagrams).
In the following we will treat the tree and loop parts of
the amplitudes separately,
\begin{equation}
A_i = A_i^{tree} + A_i^{loop} \, .
\end{equation}

\subsection{Tree Diagrams}

We expect the $s$- and $u$-channel Born diagrams (Fig. \ref{treefig} (a,b))
and the contact
diagram (Fig. \ref{treefig} (c))
to generate the structure-independent part of the amplitude
which was predicted for real Compton scattering
by Low \cite{Low} and Gell-Mann and Goldberger
\cite{GG}. In \cite{BKKM1} it has been shown that the heavy baryon
approach reproduces this LET, which must be the case for any
consistent calculation based on gauge invariance, Lorentz invariance
and crossing symmetry. For the case of VCS
we obtain
\begin{mathletters}
\begin{eqnarray}
A_1^{tree} & = & - \frac{1}{2 M} \left( 1 + \tau_3 \right)\,
,
\\
A_2^{tree} & = & \frac{1}{2 M^2} \left( 1 + \tau_3 \right)
\mid \vec q \mid \, ,
\\
A_9^{tree} & = & - \frac{1}{2 M} \left( 1 + \tau_3 \right) +
\frac{1}{2 M^2} \left( 1 + \tau_3 \right) \mid \vec q \mid \cos \theta
\nonumber\\
&+& \frac{1}{4 M^2} \left( 1 + \tau_3 \right) \frac{\mid \vec q
  \mid^2}{\omega'}
\, .
\end{eqnarray}
\end{mathletters}
Recently the LET has been extended to the case of virtual
Compton scattering on the nucleon \cite{GLT,KKS1}. We find that our results
at $O \left( p^3 \right)$ exactly agree with the predictions,
which behave as
$1/M$ and $1/M^2$ in \cite{KKS1}. However, the terms
of order $1/M^3$ which have also been derived in \cite{KKS1} as well
as the $q^2$ corrections to
the zeroeth order nucleon form factors, which are proportional to
$\frac{1}{M} \frac{1}{(4 \pi F)^2}$,\cite{BKKM1} are
beyond the predictive power of our $O \left( p^3 \right)$ chiral
calculation.\footnote{Such effects will be considered in a subsequent
${O}(p^4)$ publication.}
We note that we do not generate any contributions to the
spin-independent part of the VCS amplitude in the case of VCS on the
neutron, because the photon only couples to the nucleon charge at
tree-level in the
spin-independent sector, but not to its magnetic moment.

\subsection{Loop Diagrams}

In contrast to the three diagrams in Figs. \ref{treefig} (a) -- (c)
discussed above we expect the loop
diagrams in Fig. \ref{loopfig}
(and the $t$-channel diagram from Fig. \ref{treefig} (d))
to contribute to the ``structure-dependent,'' and
thus model-dependent terms beyond the LET. Qualitatively, this can be
understood as follows: In the sense of Low's method \cite{Low2} the
$s$- and $u$-channel diagrams of Figs. \ref{treefig} (a) and (b)
generate the most singular terms, whereas the contact interaction of
Fig. \ref{treefig} (c) is required by gauge invariance. On the other
hand, the one-particle irreducible loop diagrams of Fig. \ref{loopfig}
yield regular contributions to the VCS amplitude, which are of higher
order in the external momenta. These contributions are not predicted
by gauge invariance alone. Nevertheless, we will see that an extended
low-energy expansion
will enable us to derive {\it{some}} constraints for a calculation of the
structure-dependent amplitude (see section \ref{lowenergy}).

In the calculation of the loop contribution to the amplitudes,
$A_i^{loop}$, it turns out, as already pointed out above,
that the number of loop diagrams is reduced to 9 by the choice of the
gauge and of the velocity $v$. By power counting arguments,
the only possibility of
generating $O \left( p^3 \right)$ loop diagrams consists in using
interactions from ${\cal{L}}_{\pi N}^{(1)}$ and ${\cal{L}}_{\pi
  \pi}^{(2)}$.
As a consequence the photon-nucleon interaction
vanishes at lowest order,
${\cal{L}}^{(1)}_{\gamma N N} = 0$, and we need not consider any
loop diagrams which contain this kind of interaction.
Using the interactions of Eqs. (\ref{L1}), (\ref{L2}) and (\ref{Lpi}),
we obtain the invariant amplitudes
for the 9 diagrams in Fig. \ref{loopfig}, ${\cal{M}}_{VCS}^{(1)}$ to
${\cal{M}}_{VCS}^{(9)}$. The exact results are listed in the
appendix
.
For some diagrams we cannot carry out the integrations over
one or even two Feynman parameters analytically and could, in principle,
proceed to evaluate these integrals
numerically. However, since we want to establish a
connection with the general low-energy expansion of the
structure-dependent part of the VCS amplitude, we expand the expressions in
Eq.
(\ref{Ml}) in terms of the two external momenta, $\omega'$ and $\mid \vec q
\mid$, using Eqs. (\ref{kinex}), (\ref{q2}) and (\ref{t}).
This expansion will scale with the mass of the particle
propagating in the loop. Consequently, we will obtain a power series
in $r/m_{\pi}$ where $r$ stands for $\omega'$ or $\mid \vec q
\mid$.\footnote{This expansion does not increase the chiral order, because
$r$ and $m_{\pi}$ are both treated as ``small'' parameters.}
We could in principle
extend the $r/m_{\pi}$ expansion to an arbitrary given
order within the framework of a $O \left( p^3 \right)$ calculation. In
this context we
want to point out that it is crucial not to confuse the expansion in
$r/m_{\pi}$ with the chiral expansion in $r/M$. The result of the chiral
$O \left( p^3 \right)$ calculation, expanded
to the order $r^4$ then reads
\begin{mathletters}
\label{al}
\begin{eqnarray}
A_1^{loop} & = & \frac{g_A^2}{F^2} \frac{1}{\pi m_{\pi}} \left[
\frac{5}{96} \omega'^2 + \frac{1}{192}
\omega'
\mid \vec q \mid \cos \theta \right. \nonumber \\
 & & + \frac{17}{1920} \frac{1}{m_{\pi}^2} \omega'^4 +
 \frac{19}{1920} \frac{1}{m_{\pi}^2} \omega'^3 \mid \vec
 q \mid \cos \theta \nonumber \\
 & & - \frac{1}{384} \frac{1}{m_{\pi}^2} \omega'^2 \mid \vec q
 \mid^2 - \frac{1}{320} \frac{1}{m_{\pi}^2}
 \omega'^2 \mid \vec q \mid^2 \cos^2 \theta \nonumber \\
 & & \left. + \frac{1}{960} \frac{1}{m_{\pi}^2} \omega' \mid \vec
   q \mid^3 \cos \theta
 \right] \, , \label{a1l}
\\
A_2^{loop} & = & \frac{g_A^2}{F^2}
\frac{1}{\pi m_{\pi}}
\left[ -\frac{1}{192} \omega' \mid \vec q \mid - \frac{1}{384}
  \frac{1}{m_{\pi}^2} \omega'^3 \mid \vec q \mid\right. \nonumber\\
&+&
\left.\frac{1}{320} \frac{1}{m_{\pi}^2} \omega'^2 \mid \vec q \mid^2 \cos
\theta
- \frac{1}{960}
\frac{1}{ m_{\pi}^2} \omega' \mid \vec q \mid^3 \right] \, ,
\label{a2l}
\\
A_9^{loop} & = & \frac{g_A^2}{F^2} \frac{1}{\pi m_{\pi}}
\left[ \frac{5}{96} \omega'^2
  + \frac{17}{1920} \frac{1}{m_{\pi}^2} \omega'^4\right. \nonumber\\
&+&\left. \frac{7}{960} \frac{1}{m_{\pi}^2} \omega'^3 \mid \vec q
  \mid \cos \theta
- \frac{7}{960} \frac{1}{m_{\pi}^2}
\omega'^2 \mid \vec q \mid^2 \right]
\, . \label{a9l}
\end{eqnarray}
\end{mathletters}
In order to obtain $A_9^{loop}$, we made use
of Eqs. (\ref{ez}) and (\ref{a}).

\section{Low-Energy Expansion, Structure Coefficients and Polarizabilities}
\label{lowenergy}

In the previous section we
saw that the $O \left( p^3 \right)$ heavy baryon result
reproduces the terms required by Low's method in the tree-level
amplitudes. This section will deal with a general parametrization
of the structure- or model-dependent
part of ${\cal{M}}_{VCS}^{non-spin}$. In \cite{FS1} a general low-energy
parametrization $O$ of
the structure-dependent amplitude for virtual Compton scattering on
a spin-zero target, {\it e.g.} a pion, has been worked out.
The corresponding expression for the nucleon is more complicated, as
in Dirac space it involves a general $4 \times 4$-matrix. However, we
can apply the parametrization of \cite{FS1}
to the {\it{spin-independent}}
part of the VCS amplitude off the nucleon.
Indeed the
form of the scattering amplitude for the spin-independent part
of the reaction (\ref{process}) is
the same as for the process \cite{BT}
\begin{equation}
\gamma^* ( \varepsilon^{\mu}, q^{\mu} )
+ \pi ( p_i^{\mu} ) \rightarrow \gamma (
\varepsilon'^{\mu}, q'^{\mu} ) + \pi ( p_f^{\mu} ).
\end{equation}
Before discussing the structure-dependent terms, one has to specify
which convention has been used for splitting the total amplitude
${\cal{M}}_{VCS}$ into a Born contribution and a structure-dependent
part. This issue is addressed in \cite{KKS1,FS1} in quite some
detail. We follow the convention of \cite{GLT,KKS1} where
the Born contribution is calculated using Dirac and Pauli form
factors, $F_1$ and $F_2$. In fact, to the order considered here, the
tree-level diagrams of Fig. \ref{treefig} (a) to (c) generate the
same contributions to $A_1$, $A_2$ and $A_9$ as a covariant Born term
calculation involving $F_1$ and $F_2$ \cite{KKS1}.

The most general spin-independent contribution to the
structure-dependent VCS amplitude can then be written as
\begin{equation}
O = \varepsilon_{\mu} O^{\mu \nu}
\varepsilon'^*_{\nu}
= O^{(1)} + O^{(2)} + O^{(3)} +O^{(4)} + {{O}}(k^5) \label{lee}
\label{lexpansion},
\end{equation}
where the terms of increasing orders of $k$ ($k$ stands for $q$ or
$q'$) are the following:
\begin{mathletters}
\begin{eqnarray}
O^{(1)} & = & 0\, ,\\
O^{(2)} & = & g_0 \left[ \varepsilon \cdot q' \varepsilon'^* \cdot q - q
  \cdot q' \varepsilon \cdot \varepsilon'^* \right] \nonumber \\
 & & + \tilde{c}_1 \left[ \left( q + q' \right) \cdot \left( p_i + p_f
   \right) \left( \varepsilon \cdot \left( p_i + p_f \right)
     \varepsilon'^* \cdot q + q' \cdot \varepsilon \left( p_i + p_f
     \right)  \cdot \varepsilon'^* \right) \right. \nonumber \\
 & & \left. - 2 q \cdot q' \varepsilon
   \cdot \left( p_i + p_f \right) \varepsilon'^* \cdot \left( p_i + p_f
   \right) - 2 q \cdot \left( p_i + p_f \right) q' \cdot \left( p_i +
     p_f \right) \varepsilon \cdot \varepsilon'^* \right] \, ,
 \label{o2}
\\
O^{(3)} & = & 0\, ,
\\
O^{(4)} & = & \left[ g_{2a} q \cdot q' + g_{2b} q^2 + 4 g_{2c} q \cdot
  \left( p_i + p_f \right) q' \cdot \left( p_i + p_f \right) \right]
\left[ \varepsilon \cdot q' \varepsilon'^* \cdot q - q \cdot q'
  \varepsilon \cdot \varepsilon'^* \right] \nonumber \\
& & + \left[ {\tilde{c}}_{3a} q \cdot q' + {\tilde{c}}_{3b} q^2 + 4
  {\tilde{c}}_{3c} q \cdot \left( p_i + p_f \right) q' \cdot \left(
    p_i + p_f \right) \right] \nonumber \\
& & \times \left[ \left( q + q' \right) \cdot \left( p_i + p_f \right) \left(
\varepsilon \cdot \left( p_i + p_f \right) \varepsilon'^* \cdot q +
\varepsilon \cdot q' \varepsilon'^* \cdot \left( p_i + p_f \right)
\right) \right. \nonumber \\
& & \left. - 2 q \cdot q' \varepsilon \cdot \left( p_i + p_f \right)
\varepsilon'^* \cdot \left( p_i + p_f \right) - 2 q \cdot \left( p_i +
  p_f \right) q' \cdot \left( p_i + p_f \right) \varepsilon \cdot
\varepsilon'^* \right] \nonumber \\
& & + c_3 \left[ 2 q^2 \left( q \cdot \left( p_i + p_f \right) q'
    \cdot \left( p_i + p_f \right) \varepsilon \cdot \varepsilon'^* -
    q \cdot q' \varepsilon \cdot \left( p_i + p_f \right) \varepsilon'^*
    \cdot \left( p_i + p_f \right) \right) \right. \nonumber \\
& & \left. + \left( q + q' \right)
  \cdot \left( p_i + p_f \right) q^2 \left( \varepsilon \cdot \left(
      p_i + p_f \right) \varepsilon'^* \cdot q - \varepsilon \cdot q'
    \varepsilon'^* \cdot \left( p_i + p_f \right) \right) \right.
\nonumber \\
& & \left. + 2 \left( q + q' \right) \cdot \left( p_i + p_f \right) q \cdot
q' \varepsilon \cdot q \varepsilon'^* \cdot \left( p_i + p_f \right)
\right. \nonumber \\
& & \left.
- 4 q \cdot \left( p_i + p_f \right) q' \cdot \left( p_i + p_f \right)
\varepsilon \cdot q \varepsilon'^* \cdot q \right]
\label{o4} \, .
\end{eqnarray}
\end{mathletters}
Here we have used the same notation for the unknown structure coefficients
as in \cite{FS1}.
We then match our chiral calculation of the $O \left( p^3
\right)$ spin-independent
amplitude with the
low-energy expansion, Eq. ({\ref{lee}), demanding
\begin{equation}
- i e^2 O
= {\cal{M}}_{VCS}^{loop, non-spin}\, . \label{matching}
\end{equation}
A consistent matching procedure requires that we make use of identical
approximations for the kinematical quantities (see Eqs.\ (\ref{kinex}),
(\ref{q2}) and (\ref{t})) in both the chiral loop calculation and the
low-energy expansion, Eq.\ (\ref{lee}). To be specific, we have neglected
corrections of the type $\omega'/M$ and $\mid \vec q \mid / M$ (and
higher).
After this procedure we
are able to determine the unknown structure coefficients in
Eqs.\ (\ref{o2}) and (\ref{o4}) from our chiral calculation.
Using Eq.\ (\ref{lee}) we obtain the following
parametrization of the spin-independent amplitudes:
\begin{mathletters}
\label{ale}
\begin{eqnarray}
A_1^{structure} & = & \left. \omega'^2 \left[
    - g_0 - 8 M^2 {\tilde{c}}_1 \right] + \omega' \mid \vec q \mid \cos
  \theta g_0
\right.
\nonumber
\\
& & \left.
+ \omega'^4 \left[ - g_{2a} - g_{2b} - 16 M^2 g_{2c} + 8 M^2 c_3 - 8
  M^2 {\tilde{c}}_{3a} - 8 M^2
  {\tilde{c}}_{3b} - 128 M^4 {\tilde{c}}_{3c} \right]
\right. \nonumber \\
& & \left. + \omega'^3 \mid
\vec q \mid  \cos \theta \left[ 2 g_{2a} + g_{2b} + 16 M^2 g_{2c} +
  8 M^2 {\tilde{c}}_{3a} \right] \right. \nonumber \\
& & \left. + \omega'^2 \mid \vec q \mid^2 \left[
  g_{2b} - 8 M^2 c_3 + 8 M^2 {\tilde{c}}_{3b} \right]
+ \omega'^2 \mid \vec q \mid^2 \cos^2 \theta \left[ - g_{2a} \right]
\right. \nonumber
\\
& & \left. + \omega' \mid \vec q \mid^3 \cos \theta \left[ - g_{2b} \right]
\right. \, ,\label{a1le}
\\
A_2^{structure} & = & \left. \omega' \mid \vec q \mid \left[ - g_0
  \right]
\right. \nonumber
\\
& & \left. +
  \omega'^3 \mid \vec q \mid
\left[ - g_{2a} - g_{2b} - 16 M^2 g_{2c} \right] + \omega'^2
  \mid \vec q \mid^2 \cos \theta g_{2a}
+ \omega' \mid \vec q
  \mid^3 g_{2b} \right. \, , \label{a2le}
\\
A_9^{structure} & = & \left. \omega'^2 \left[ - g_0 - 8 M^2 {\tilde{c}}_1
  \right] \right. \nonumber \\
& & \left. + \omega'^4 \left[ - g_{2a} - g_{2b} - 16 M^2 g_{2c} + 8 M^2
    c_3 - 8 M^2 {\tilde{c}}_{3a}- 8 M^2 {\tilde{c}}_{3b} - 128 M^4
{\tilde{c}}_{3c} \right] \right. \nonumber \\
& & \left. + \omega'^3 \mid \vec q \mid \cos \theta \left[ g_{2a} + 8 M^2
  {\tilde{c}}_{3a} \right] + \omega'^2 \mid \vec q \mid^2 \left[
  g_{2b} + 8 M^2 c_3 + 8 M^2 {\tilde{c}}_{3b} \right]
\right. \, \label{a9le}.
\quad
\end{eqnarray}
\end{mathletters}
Demanding the validity of the matching relation Eq. (\ref{matching})
we obtain a system of
linear equations for the structure coefficients, yielding
\begin{mathletters}
\begin{eqnarray}
g_0 & = & \frac{1}{192} \frac{g_A^2}{F^2} \frac{1}{\pi m_{\pi}}
\, ,
\\
{\tilde{c}}_1 & = & - \frac{11}{192} \frac{1}{8 M^2} \frac{
g_A^2}{F^2}
\frac{1}{\pi m_{\pi}}\, ,
\\
g_{2a} & = & \frac{1}{320} \frac{g_A^2}{F^2} \frac{1}{\pi
m_{\pi}^3}\, ,
\\
g_{2b} & = & - \frac{1}{960} \frac{g_A^2}{F^2} \frac{1}{\pi m_{\pi}^3}
\,
\\
g_{2c} & = & \frac{1}{1920} \frac{1}{16 M^2}
\frac{g_A^2}{F^2} \frac{1}{\pi m_{\pi}^3}\, ,
\\
c_3 & = & - \frac{3}{1280} \frac{1}{8 M^2}
\frac{g_A^2}{F^2} \frac{1}{\pi m_{\pi}^3}\, ,
\\
{\tilde{c}}_{3a} & = & \frac{1}{240} \frac{1}{8 M^2}
\frac{g_A^2}{F^2} \frac{1}{\pi m_{\pi}^3}\, ,
\\
{\tilde{c}}_{3b} & = & - \frac{1}{256} \frac{1}{8 M^2}
\frac{g_A^2}{F^2} \frac{1}{\pi m_{\pi}^3}\, , \\
{\tilde{c}}_{3c} & = &
-
\frac{9}{640} \frac{1}{128 M^4} \frac{g_A^2}{F^2} \frac{1}{\pi m_{\pi}^3}\, .
\end{eqnarray}
\end{mathletters}
Moreover, we wish to stress that it is actually a {\it{prediction}} of
the low-energy expansion, Eq.\ (\ref{a1le}), that the expression for $A_1$
{\it cannot} involve the structures $\mid \vec q
\mid^2$ and $\mid \vec q \mid^4$.
Similarly the expansion of the amplitude $A_9$, Eq. (\ref{a9le}),
excludes structures
of the type $\omega' \mid \vec q \mid \cos \theta$,
$\omega'^2 \mid \vec q \mid^2 \cos^2 \theta$, $\omega' \mid \vec q
\mid^3 \cos \theta$ and $\mid \vec q \mid^4$.
When we compare these predictions with the result of our loop
calculation, we find that each of these constraints is satisfied.
We also note that our $O \left( p^3 \right)$ calculation gives a zero
result for structures with an odd power of $r \,\,\,(r \in
\left\{\omega', \mid \vec q \mid \right\})$, which is
required by the general expression in \cite{FS1}
obtained by imposing gauge invariance, crossing symmetry and the
discrete symmetries.

It is useful at this point to interpret the structure coefficients
which we have obtained. The coefficients which originate from
$O^{(2)}$ --- $g_0$ and ${\tilde{c}}_1$ ---
are closely related to the electric and
magnetic polarizabilities of the nucleon in real Compton scattering,
${{\alpha}}_0$ and ${{\beta}}_0$. This
can be seen by considering
the limit of real Compton scattering in the
c.m.\ system, {\it i.e.}
$\mid \vec q \mid = \omega = \omega'$. If we apply the definitions for
the polarizabilities,
(see, {\it e.g.}, \cite{BKM1})
\begin{eqnarray}
{{\alpha}}_0 + {{\beta}}_0 & = &
\frac{e^2}{8 \pi} \frac{\partial^2}{\partial
  \omega^2}  A_1 \left(\omega=0, \theta=0 \right) \, , \nonumber\\
{{\beta}}_0 & = & - \frac{e^2}{4 \pi} \left( \frac{A_2}{\omega' \mid
    \vec q \mid} \right) \left(\omega=0, \theta=0
\right) \, ,
\end{eqnarray}
to Eqs. (\ref{a1l}) and (\ref{a2l}), we determine
\begin{eqnarray}
{{\alpha}}_0 & = & \frac{5 e^2 g_A^2}{384 \pi^2 m_{\pi} F^2}\, , \nonumber\\
{{\beta}}_0 & = & \frac{e^2 g_A^2}{768 \pi^2 m_{\pi} F^2}  \, ,
\end{eqnarray}
which coincide with the results found in \cite{BKKM1}. We can then express
the structure constants $g_0$ and ${\tilde{c}}_1$ in terms of
${{\alpha}}_0$ and ${{\beta}}_0$,
\begin{mathletters}
\begin{eqnarray}
g_0 & = & \frac{4 \pi}{e^2} {{\beta}}_0\, , \label{g_0} \\
{\tilde{c}}_1 & = & - \frac{\pi}{2 e^2 M^2} \left( {{\alpha}}_0 +
{{\beta}}_0 \right) \label{c_1}\, .
\end{eqnarray}
\end{mathletters}
An alternative way by which
to derive this relation without explicitly taking the real
Compton limit is to compare our results for $A_1$, $A_2$ and $A_9$
with the low-energy expansion of the VCS amplitude of \cite{KKS1}. We
find that this method yields identical results for the
polarizabilities ${{\alpha}}_0$ and ${{\beta}}_0$ once the
transformation to our convention is made.

\section{Comparison with Other Calculations}

We now investigate the connection of our results
with the multipole expansion of \cite{GLT}, where possible.
We begin by recalling the primary features of such an expansion.  Guichon, Liu
and Thomas parametrize the structure-dependent part of the
VCS amplitude in terms of essentially the same kinematical
quantities as do we, namely, $\omega'$, $\mid \vec q \mid$, $\theta$.
However, they expand the
structure-dependent part of the amplitude
in terms of $\omega'$, {\it keeping only the first
non-vanishing, linear term.}  (They do not consider
$\omega'^2$ terms because they are of higher order in their
expansion.)  The multipole
expansion then generates various combinations of $\cos \theta$ and powers
of $\mid \vec q \mid$ and, furthermore, suggests the definition of
generalized polarizabilities characterized by angular momentum quantum
numbers of the respective partial waves. The crucial difference in
comparison with the
low-energy expansion introduced above consists in the feature that the
starting point of the Guichon {\it et al.} analysis is the $\omega'$ expansion around
$\omega'=0$, wherein $\mid \vec q \mid$ can be chosen
arbitrarily, ({\it i.e.} not necessarily small), once $\omega'$ has a small
value. One can see the difference between the
two expansion schemes most easily by looking
at the quantity $\omega$. Rewriting Eq.\ (\ref{kinex}), we obtain
\begin{equation}
\omega = \omega' + \frac{\omega'^2}{2 M} - \frac{\mid \vec q \mid^2}{2
  M} + {{O}}\left(\frac{r^4}{M^3} \right) \, . \label{omchi}
\end{equation}
On the other hand, Guichon, Liu and Thomas use
\begin{equation}
\omega \mid_{\omega'=0}
= - \frac{\mid \vec q \mid^2}{2 M} + {{O}}\left(\frac{\mid \vec q
  \mid^4}{M^3} \right)\, . \label{ommul}
\end{equation}
From Eqs. (\ref{omchi}) and (\ref{ommul}) we conclude that we
will not, in general, be able to express the results of our calculation for
the structure coefficients in terms of the generalized
polarizabilities.
In our approximation scheme we take the quantities
$\omega$ and $\omega'$ to be equal and consider terms up to
$\omega'^4$ whereas \cite{GLT} only keeps one power of $\omega'$ and
starts the expansion of $\omega$ with a term which due to its $1/M$
suppression would be higher
order in our scheme.
Using Eq.\ (\ref{ommul})
the expansions of the model-dependent parts of the spin-independent
amplitudes in \cite{GLT} read
\begin{mathletters}
\label{anb}
\begin{eqnarray}
A_1^{non-Born} & = & - \sqrt{\frac{3}{8}} \omega' \mid \vec q \mid \cos
\theta P^{(11,11)0} \left(\mid \vec q \mid \right) \nonumber\\
&-& \sqrt{\frac{3}{2}} \omega'
\omega\!\mid_{\omega'=0} P^{(01,01)0} \left(\mid \vec q \mid \right)
\nonumber \\
& -& \frac{3}{2} \omega' \mid \vec
q \mid^2 {\hat{P}}^{(01,1)0} \left(\mid \vec q \mid \right)
+ {{O}} \left( \omega'^2 \right) \, , \label{a1nb}
\\
A_2^{non-Born} & = & \sqrt{\frac{3}{8}} \omega' \mid \vec q \mid
P^{(11,11)0} \left(\mid \vec q \mid \right)
+ {{O}} \left( \omega'^2 \right) \, , \label{a2nb}
\\
A_9^{non-Born} & = & - \sqrt{\frac{3}{2}} \omega'
\omega \mid_{\omega'=0}
P^{(01,01)0} \left(\mid \vec q \mid \right) +
{{O}} \left( \omega'^2 \right)\label{a9nb}\, ,
\end{eqnarray}
\end{mathletters}
where we had to modify the definition of $A_9^{non-Born}$ in
\cite{GLT}, because therein the Lorentz gauge, $\varepsilon \cdot q =
0$, is used, while we work in the gauge defined in Eq.\ (\ref{a}).
Note that the generalized polarizabilities $P^{(11,11)0}$, $P^{(01,01)0}$  and
${\hat{P}}^{(01,1)0}$ are functions of $\mid \vec q \mid$ only.

Let us now compare Eq.\ (\ref{a1l}) with Eq.\ (\ref{a1nb}). First of all,
we note that in virtual Compton scattering the term containing
$\alpha_0$ is proportional to $\omega \omega'$ before we have made any
kinematical approximations. This can be seen from Eqs.\ (\ref{o2}) and
(\ref{c_1}). When we insert Eq.\ (\ref{omchi}) for $\omega$ the leading
term will be proportional to $\omega'^2$, whereas applying Eq.\ (\ref{ommul})
the leading term will be proportional to $\omega' \omega
\mid_{\omega'=0}$. The important observation here is that in both
expansion schemes these terms have the same coefficient which is
proportional to $\alpha_0$. We conclude that our result for the
$\omega'^2$ term in Eq.\ (\ref{a1l})
serves as a prediction for $\alpha_0$ in \cite{GLT},
\begin{equation}
P^{(01,01)0} \left( \mid \vec q \mid = 0 \right)
= - \frac{4 \pi}{e^2} \sqrt{\frac{2}{3}} \alpha_0
= \sqrt{\frac{2}{3}} \left( g_0 + 8 M^2 {\tilde{c}}_1 \right)
\, . \label{alphanb}
\end{equation}
With an analogous chain of arguments we find that the term $\omega' \mid
\vec q \mid \cos \theta$ in Eq.\ (\ref{a1nb})
can be directly compared with the corresponding term in
Eq.\ (\ref{a1l}), yielding
\begin{equation}
P^{(11,11)0} \left( \mid \vec q \mid = 0 \right)
= - \frac{4 \pi}{e^2} \sqrt{\frac{8}{3}} \beta_0
= - \sqrt{\frac{8}{3}} g_0
\, .
\end{equation}
We see that the result for the magnetic polarizability $\beta_0$
 obtained from Eqs.\ (\ref{a1l}) and (\ref{a1nb})
is consistent with the fact that both Eq.\ (\ref{a2l}) and Eq.\ (\ref{a2nb})
can also be parametrized in terms of this polarizability as a
pre-factor of the $\omega' \mid \vec q \mid$ term. As regards
Eq.\ (\ref{a9nb}) we observe that the parametrization
in terms of $P^{(01,01)0}$
is consistent with our result for the pre-factor of the $\omega' \mid
\vec q \mid$ term in Eq.\ (\ref{a9l}). Now what about
${\hat{P}}^{(01,1)0} \left( \mid \vec q \mid  = 0 \right)$
in Eq.\ (\ref{a1nb})? In order to determine this polarizability from our
calculation, one would have to calculate the coefficient of the
$\omega' \mid \vec q \mid^2$ term in Eq.\ (\ref{a1l}). Due to the
approximations of Eqs.\ (\ref{kinex}), (\ref{q2}) and (\ref{t}),
a $O \left(p^3 \right)$
heavy baryon calculation does not generate a term of this type. So we
cannot determine this polarizability in our calculation. We expect the
leading term contributing to
${\hat{P}}^{\left(01,1\right)0} \left( \mid \vec q \mid =
  0 \right)$ to be at least $1/M$ suppressed.  Hence its evaluation
requires a consistent $O\left( p^4 \right)$ calculation, which is in progress.
This is borne out by the linear sigma model calculation
\cite{MD} which finds that the coefficient of $\omega' \mid \vec q \mid^2$
is indeed $1/M$ suppressed. Furthermore, in
\cite{MD} it was shown numerically that ${\hat{P}}^{(01,1)0}$ is
not independent of the generalized polarizabilities $\alpha$ and $\beta$.

Let us turn to the remaining seven structure constants,
$g_{2a}$, $g_{2b}$, $g_{2c}$, $c_3$, ${\tilde{c}}_{3a}$,
${\tilde{c}}_{3b}$, ${\tilde{c}}_{3c}$, which parametrize the $r^4$
terms in Eq.\ (\ref{ale}).
First we observe in Eqs.\ (\ref{a1l}) and (\ref{a1nb}) that, strictly
speaking, we
will only be able to compare terms which contain $\omega'$
(and under certain circumstances also $\omega'^2$)
but not the terms with $\omega'^3$ and $\omega'^4$. These
are higher-order corrections in Eq.\ (\ref{a1nb}) but not in
Eq.\ (\ref{a1l}), where $\omega'$ and $\mid \vec q \mid$ are counted as
being of the
same order. This suggests that the multipole expansion of Eq.\
(\ref{anb}) and the chiral expansion
of Eq.\ (\ref{al}) have different domains of
application. We will discuss the appropriate kinematical region for each
of the two expansions later. We conclude that the terms
$\omega'^3, \omega'^4$ belong to the model-dependent piece of the VCS
amplitude, but do not fit into the expansion scheme in terms of
the generalized polarizabilities of Eq.\ (\ref{anb}).
Let us now discuss the $\omega'^2 \mid \vec q \mid^2 \cos^2
\theta$ term in Eq.\ (\ref{a1l}). We cannot expect to find an analogous
term in Eq.\ (\ref{a1nb}), because the quadratic dependence on $\cos
\theta$ indicates a term of higher multipolarity which has not been
considered in the special application of Eq.\ (\ref{a1nb}), where the angular
momentum of the final state photon is fixed at $L' = 1 $.
As regards the $\omega'^2 \mid \vec q \mid^2$ terms in Eqs.\ (\ref{a1l}) and
(\ref{a9l}) we find that the factors $\omega'^2$ originate from the
product $\omega \omega'$ before we apply Eq.\ (\ref{kinex}). This can be
seen from Eqs.\ (\ref{o4}), (\ref{a1le}) and (\ref{a9le}). As a result we
might be tempted to
interpret the coefficient of the $\omega'^2 \mid \vec q \mid^2$
terms as $\frac{d}{d \mid \vec q \mid^2} \alpha \left( \mid \vec q
  \mid = 0 \right)$. However, we find different solutions from matching the
transverse and the longitudinal results of Eqs.\ (\ref{a1l}) and
(\ref{a9l})
with
Eqs.\ (\ref{a1nb}) and (\ref{a9nb}). Inspecting Eqs.\ (\ref{a1le}) and
(\ref{a9le}) we observe that the difference between the transverse and
the longitudinal result can be explained by the fact that the
structure constant $c_3$ enters the terms with a different sign.
If we recall that the transverse amplitude $A_1^{loop}$ contains
contributions from higher multipoles than $L'=1$, we come to the
conclusion that the $\omega'^2 \mid \vec q \mid^2$ term can also get
contributions from such multipoles, in particular from the constant
part of the Legendre polynomial for $L' = 2 $.
For this reason we cannot
determine $\frac{d}{d \mid \vec q \mid^2} \alpha \left( \mid \vec q
  \mid = 0 \right)$ from this amplitude. In the longitudinal amplitude
$A_9^{loop}$, however, we find no contributions from higher
multipoles. For this reason we use the longitudinal amplitude for the
determination of $\frac{d}{d \mid \vec q \mid^2} \alpha \left( \mid \vec q
  \mid = 0 \right)$ and, using Eqs.\
(\ref{a9l}), (\ref{a9le}) and (\ref{a9nb}),
arrive at
\begin{eqnarray}
\frac{d}{d \mid \vec q \mid^2} \alpha \left( \mid \vec q
  \mid\right. &=&\left. 0 \right)
= \frac{e^2}{4 \pi} \left( g_{2b} + 8 M^2 c_3 + 8 M^2 {\tilde{c_{3b}}}
  \right) \nonumber\\
 &=& - \frac{7 e^2 g_A^2}{3840 \pi^2 m_{\pi} F^2} \, .
\end{eqnarray}
The situation is more obvious for $\frac{d}{d \mid \vec q \mid^2}
\beta \left( \mid \vec q
  \mid = 0\right)$. If we compare
the term $\omega' \mid \vec q \mid^3 \cos \theta$ in Eq.\ (\ref{a1l})
with Eqs.\ (\ref{a1le}) and (\ref{a1nb}) we find
\begin{equation}
\frac{d}{d \mid \vec q \mid^2} \beta \left( \mid \vec q
  \mid = 0 \right)  = -\frac{e^2}{4 \pi} g_{2b}    =
 \frac{e^2 g_A^2}{3840 \pi^2 m_{\pi} F^2}
\, .
\end{equation}
From Eq.\ (\ref{a2nb}) we can read off that we could expect the same
coefficient for the $\omega \mid \vec q \mid^3$ term in Eqs.\ (\ref{a2l})
and (\ref{a2le}), and we find that our results are consistent
with Eq.\ (\ref{a2nb}).  This is supported by the fact that we have no
indication of higher multipoles entering $A_2^{loop}$ in our
calculation which could obscure the $L'=1$ results.

Summing up, we have established a connection between the
multipole expansion \cite{GLT} and the low-energy coefficients
\cite{FS1}
$g_0$, ${\tilde{c}}_1$, $g_{2b}$ and the linear combination
$c_3 + {\tilde{c}}_{3b}$,
which is reflected in the following expansion of the electric
and magnetic polarizabilities of the nucleon,
\begin{mathletters}
\begin{eqnarray}
\alpha \left( \mid \vec q \mid \right) & = & \alpha_0 \left( 1 -
  \frac{7}{50} \frac{\mid \vec q \mid^2}{m_{\pi}^2} + {{O}} \left(
    \frac{\mid \vec q \mid^4}{m_{\pi}^4} \right)  \right) \nonumber \\
& = &
\frac{e^2}{4 \pi} \left( - \left[ g_0 + 8 M^2 {\tilde{c}}_1 \right]
\right.\nonumber\\
&+& \left.\mid \vec q
  \mid^2 \left( g_{2b} + 8 M^2 c_3 + 8 M^2 {\tilde{c}}_{3b} \right)
+ {{O}} \left(
    \frac{\mid \vec q \mid^4}{m_{\pi}^4} \right)
\right)
\, ,
\label{alphaex} \\
\beta \left( \mid \vec q \mid \right) & = & \beta_0 \left( 1 +
  \frac{1}{5} \frac{\mid \vec q \mid^2}{m_{\pi}^2} + {{O}} \left(
    \frac{\mid \vec q \mid^4}{m_{\pi}^4} \right) \right) \nonumber \\
& = & \frac{e^2}{4 \pi} \left( g_0 - \mid \vec q \mid^2 g_{2b}
+ {{O}} \left(
    \frac{\mid \vec q \mid^4}{m_{\pi}^4} \right)
\right)
\, . \label{betaex}
\end{eqnarray}
\end{mathletters}
We wish to point out
Eqs.\ (\ref{alphaex})
and (\ref{betaex}) exactly agree with the corresponding terms in the
relativistic field-theoretical calculation within the linear sigma
model \cite{MD}. In comparison with the results of \cite{GLT,Van,GLT2}
our value for the electric polarizability of the proton,
$\alpha_0=12.8 \times 10^{-4} \, \mathrm{fm}^3$ is
much larger and our value for the magnetic polarizability of the
proton,
$\beta_0=1.3 \times 10^{-4} \, \mathrm{fm}^3$ is smaller than in those
calculations. (We have used the numerical values $m_{\pi} = 135 \,
\mathrm{MeV}$, $F = 92.4 \, \mathrm{MeV}$ and $g_A = 1.26$.)
The slope of the corresponding generalized electric polarizability,
$\frac{d}{d \mid \vec q \mid^2} \alpha \left( \mid \vec q \mid = 0
\right)$, is found to be considerably larger than in the effective
Lagrangian \cite{Van} and the constituent quark model \cite{GLT,GLT2}
calculations.  The slope of the generalized magnetic polarizability,
$\frac{d}{d \mid \vec q \mid^2} \beta \left( \mid \vec q \mid = 0
\right)$, even has a different sign compared with other calculations.
For the neutron we find the same analytical expressions
for these quantities as for the proton.

We cannot give an interpretation of the structure coefficients
$g_{2a}$, $g_{2c}$, ${\tilde{c}}_{3a}$, ${\tilde{c}}_{3c}$ and the
missing linear combination $c_3 - {\tilde{c}}_{3b}$
in terms of generalized polarizabilities. However, all structure
coefficients in Eq.\ (\ref{o4}), not only the subset which we could
interpret in terms of generalized polarizabilities, have to be
considered if one performs an experiment in a kinematical region where
$\omega'$ is comparable with $\mid \vec q \mid$. Furthermore, we find
that, in addition to the convergence radius of the chiral Lagrangian,
which is given by $r/M$, the inelastic threshold of single pion
production sets a primary limit to our calculation. The low-energy
parametrization in terms of structure coefficients is only valid in a
regime below the pion threshold, where $\omega$ and $\mid \vec q \mid$ do
not approach $m_{\pi}$.
This is clearly a different kinematical regime than the one of the
multipole expansion \cite{GLT}, which is predominantly suited for
arbitrary $\mid \vec q \mid$, which are much larger than the very small
$\omega'$. (Of course, in principle our calculation is exact for
$\omega' < m_{\pi}$ and $\mid \vec q \mid < m_{\pi}$ if we keep the
finite integrals in Eq.\ (\ref{Ml}) instead of expanding
the integrands, as done above.)

In conclusion, we have investigated the spin-averaged amplitude of
virtual Compton scattering within the framework of heavy baryon chiral
perturbation theory to $O \left( p^3 \right)$. We have determined the
generalized spin-independent electromagnetic polarizabilities of the
nucleon \cite{GLT}. In order to obtain analytic expressions,
we have restricted our calculation
to $\alpha_0$ and $\beta_0$ of real Compton scattering
and the
slopes of the generalized polarizabilities with respect to $\mid \vec
q \mid^2$. We have also performed an alternative low-energy expansion
of the VCS amplitude \cite{FS1} which is not restricted to first order in the
energy of the outgoing photon as is the multipole expansion. We made a
prediction, based on chiral symmetry, for the 9 {\it a priori} unknown
structure coefficients characterizing the spin-independent part of the
VCS amplitude up to ${O}(r^4)$. All predictions of the loop calculation
are the same for the proton and the neutron at ${O}(p^3)$, but
will differ in a ${O}(p^4)$ evaluation.

\appendix
\section{Loop Amplitudes}
\label{appendix}
Using standard techniques we obtain the following invariant amplitudes
for the Feynman diagrams in Fig. \ref{loopfig} with the Lagrangian
interactions, Eqs.\ ($\ref{L1}$), ($\ref{L2}$) and ($\ref{Lpi}$):
\begin{mathletters}
\begin{eqnarray}
{\cal{M}}_{VCS}^{(1)} & = &
- \frac{e^2 g_A^2}{2 F^2}
\int \frac{d^d l}{\left(2 \pi \right)^d}
{\bar{N_v}}
\varepsilon'^* \cdot a
\frac{1}{v \cdot \left(l+t_i+q \right)+ i 0^+ }
\frac{1}{l^2 - m_{\pi}^2 + i 0^+}
N_v,\\
{\cal{M}}_{VCS}^{(2)} & = &
- \frac{e^2 g_A^2}{2 F^2}
\int \frac{d^d l}{\left(2 \pi \right)^d}
{\bar{N_v}}
\varepsilon'^* \cdot a
\frac{1}{v \cdot \left(l+t_i-q' \right)+ i 0^+ }
\frac{1}{l^2 - m_{\pi}^2 + i 0^+}
N_v,\\
{\cal{M}}_{VCS}^{(3)} & = &
\frac{e^2 g_A^2}{2 F^2}
\int \frac{d^d l}{\left(2 \pi \right)^d}
{\bar{N_v}}
a \cdot l
{\varepsilon'^*} \cdot \left(2 l - q' \right)
\nonumber \\
& & \times
\frac{1}{v \cdot \left(l+t_i+q-q' \right)+ i 0^+ }
\frac{1}{l^2 - m_{\pi}^2 + i 0^+}
\frac{1}{\left( l - q' \right)^2 - m_{\pi}^2 + i 0^+}
N_v,\\
{\cal{M}}_{VCS}^{(4)} & = &
\frac{e^2 g_A^2}{2 F^2}
\int \frac{d^d l}{\left(2 \pi \right)^d}
{\bar{N_v}}
{\varepsilon'^*} \cdot l
a \cdot \left(2 l + q \right)
\nonumber \\
& & \times
\frac{1}{v \cdot \left(l+t_i+q-q' \right) + i 0^+}
\frac{1}{l^2 - m_{\pi}^2 + i 0^+ }
\frac{1}{\left( l + q \right)^2 - m_{\pi}^2 + i 0^+ }
N_v,\\
{\cal{M}}_{VCS}^{(5)} & = &
\frac{e^2 g_A^2}{2 F^2}
\int \frac{d^d l}{\left(2 \pi \right)^d}
{\bar{N_v}}
{\varepsilon'^*} \cdot l
a \cdot \left(2 l - q \right)
\nonumber \\
& & \times
\frac{1}{v \cdot \left(l+t_i \right) + i 0^+ }
\frac{1}{l^2 - m_{\pi}^2 + i 0^+ }
\frac{1}{\left( l - q \right)^2 - m_{\pi}^2 + i 0^+ }
N_v,\\
{\cal{M}}_{VCS}^{(6)} & = &
\frac{e^2 g_A^2}{2 F^2}
\int \frac{d^d l}{\left(2 \pi \right)^d}
{\bar{N_v}}
{a} \cdot l
{\varepsilon'^*} \cdot \left(2 l + q' \right)
\nonumber \\
& & \times
\frac{1}{v \cdot \left(l+t_i \right) + i 0^+ }
\frac{1}{l^2 - m_{\pi}^2 + i 0^+ }
\frac{1}{\left( l + q' \right)^2 - m_{\pi}^2 + i 0^+ }
N_v,\\
{\cal{M}}_{VCS}^{(7)} & = &
\frac{e^2 g_A^2}{2 F^2}
\int \frac{d^d l}{\left(2 \pi \right)^d}
{\bar{N_v}}
\left(
v \cdot \left(l+q'-q \right) v \cdot l
- l \cdot \left( l + q' - q \right)
\right)
\nonumber \\
& & \times
{\varepsilon'^*} \cdot \left(2 \left( l- q \right)  + q' \right)
a \cdot \left( 2 l - q \right)
\nonumber \\
& & \times
\frac{1}{v \cdot \left(l+t_i \right)+ i 0^+ }
\frac{1}{l^2 - m_{\pi}^2 + i 0^+ }
\frac{1}{\left( l - q \right)^2 - m_{\pi}^2 + i 0^+ }
\frac{1}{\left( l + q' - q\right)^2 - m_{\pi}^2 + i 0^+ }
N_v,\\
{\cal{M}}_{VCS}^{(8)} & = &
\frac{e^2 g_A^2}{2 F^2}
\int \frac{d^d l}{\left(2 \pi \right)^d}
{\bar{N_v}}
\left(
v \cdot \left(l+q'-q \right) v \cdot l
- l \cdot \left( l + q' - q \right)
\right)
\nonumber \\
& & \times
a \cdot \left(2 \left( l+ q' \right)  - q \right)
{\varepsilon'^*} \cdot \left( 2 l + q' \right)
\nonumber \\
& & \times
\frac{1}{v \cdot \left(l+t_i \right) + i 0^+ }
\frac{1}{l^2 - m_{\pi}^2 + i 0^+ }
\frac{1}{\left( l + q' \right)^2 - m_{\pi}^2 + i 0^+ }
\frac{1}{\left( l + q' - q\right)^2 - m_{\pi}^2 + i 0^+ }
N_v,\\
{\cal{M}}_{VCS}^{(9)} & = &
- \frac{e^2 g_A^2}{F^2}
\int \frac{d^d l}{\left(2 \pi \right)^d}
{\bar{N_v}}
\left(
v \cdot \left(l+q'-q \right) v \cdot l
- l \cdot \left( l + q' - q \right)
\right)
\nonumber \\
& & \times
{\varepsilon'^*} \cdot a
\frac{1}{v \cdot \left(l+t_i \right) + i 0^+ }
\frac{1}{l^2 - m_{\pi}^2 + i 0^+ }
\frac{1}{\left( l + q' - q\right)^2 - m_{\pi}^2 + i 0^+ }
N_v \, .
\end{eqnarray}
\end{mathletters}
We have only retained the spin-independent parts of the
$d$-dimensional integrals. We note that we do not find any dependence
on the nucleon isospin, which means that the loop contributions for
proton and neutron are identical.
Carrying out the integration over the loop momentum $l$ we find that
the sum of all amplitudes is finite. The individual amplitudes 
can be written as
\begin{mathletters}
\label{Ml}
\begin{eqnarray}
{\cal{M}}_{VCS}^{(1)} & = & - \frac{i}{2} \frac{e^2 g_A^2}{F^2}
a \cdot \varepsilon'^* J_0 \left(
  \omega, m_{\pi}^2 \right) \, , \label{M_1l} \\
{\cal{M}}_{VCS}^{(2)} & = & - \frac{i}{2} \frac{e^2 g_A^2}{F^2}
a \cdot \varepsilon'^* J_0 \left(
 - \omega', m_{\pi}^2 \right) \, , \\
{\cal{M}}_{VCS}^{(3)} & = & \frac{i e^2 g_A^2}{F^2} a
\cdot \varepsilon'^* \int_0^1 d x J_2' \left(
  - \omega' \left( 1 -x \right) + \omega, m_{\pi}^2 \right) \, , \\
{\cal{M}}_{VCS}^{(4)} & = & \frac{i e^2 g_A^2}{F^2} \left[ a \cdot
\varepsilon'^* \int_0^1 d x J_2' \left(
  - \omega' + \omega \left( 1 -x \right), m_{\pi}^2 - q^2 x \left( 1 - x
  \right)
\right) \right. \nonumber \\
& & \left. + \varepsilon'^* \cdot q a \cdot q \int_0^1 d x
\frac{x}{2} \left( 2 x - 1 \right) J_0' \left( - \omega' + \omega \left(
    1 - x \right), m_{\pi}^2 - q^2 x \left( 1 - x \right) \right)
\right]
\, , \\
{\cal{M}}_{VCS}^{(5)} & = &  \frac{i e^2 g_A^2}{F^2} \left[ a \cdot
\varepsilon'^* \int_0^1 d x J_2' \left(
  \omega x, m_{\pi}^2 - q^2 x \left( 1 - x \right) \right)
\right. \nonumber \\
& & \left. +  \varepsilon'^* \cdot q a \cdot q \int_0^1 d x
\frac{x}{2} \left( 2 x - 1 \right) J_0' \left( \omega x, m_{\pi}^2
- q^2 x \left( 1 - x \right) \right)
\right]\, , \\
{\cal{M}}_{VCS}^{(6)} & = & \frac{i e^2 g_A^2}{F^2} a
\cdot \varepsilon'^* \int_0^1 d x J_2' \left(
  -\omega' x , m_{\pi}^2 \right) \, , \\
{\cal{M}}_{VCS}^{(7)} & = & 2 \frac{i e^2 g_A^2}{F^2}
\int_0^1 d x \int_0^1 dy \nonumber \\
& & \times \left[ \vphantom{\frac{1}{2}}
a \cdot \varepsilon'^*
\left[ \vphantom{\frac{1}{2}}
- 5 \left( 1 - y \right) \right. \right. \nonumber \\
& & \left. \left. \times J_6'' \left( \omega \left(y + x
      \left( 1 - y \right) \right) - \omega' y, m_{\pi}^2
- t \left( 1 - x \right) y \left(
      1 - y \right) - q^2 x \left( 1 - x \right)
    \left( 1 - y \right)^2 \right) \right. \right. \nonumber \\
& & \left. \left. + \left( \frac{1}{2} \left( t - q^2 \right) \left( 1 - y
    \right)^2 \left( y + x \left( 1 - y \right) + y \left( 1 - x
      \right) \right)
\right. \right. \right. \nonumber \\
& & \left. \left. \left.
+ q^2 \left( 1 - x
    \right)
\left( 1 - y \right)^2 \left( y + x \left( 1 - y \right) \right) - \omega'^2
    y \left( 1 - y \right)^2
\right. \right. \right. \nonumber \\
& & \left. \left. \left. - \omega^2 \left(
      1 - x \right)
\left( 1 - y \right)^2 \left( y + x \left( 1 - y \right) \right) + \omega
    \omega' \left( 1 - y \right)^2 \left( y + x \left( 1 - y \right) +
      y \left( 1 - x \right) \right) \vphantom{\frac{1}{2}}
\right) \right. \right. \nonumber \\
& & \left. \left. \times
J_2'' \left( \omega \left( y + x \left( 1 - y \right) \right) -
  \omega' y, m_{\pi}^2 - t \left( 1 - x \right) y \left( 1 - y \right)
  - q^2 x \left( 1-x \right) \left( 1 - y \right)^2 \right)
\vphantom{\frac{1}{2}}
\right]
\right. \nonumber \\
& & \left. + a \cdot q' \varepsilon'^* \cdot q
\left[ \vphantom{\frac{1}{2}}
\left( \left( 1 - x \right) \left( 1 - y \right)^2 \left(
      1 - 6 y \right) +
y \left( 1 - y
    \right) \left( x + y \left( 1 - x \right) \right)
\right) \right. \right. \nonumber \\
& & \left. \times J_2'' \left( \omega \left( y + x \left( 1 - y
      \right) \right) - \omega' y, m_{\pi}^2 - t \left( 1 - x \right)
    y \left( 1 - y
    \right) - q^2 x \left( 1 - x \right) \left( 1
      - y \right)^2 \right) \right.
\nonumber \\
& & \left. + \left( \vphantom{\frac{1}{2}}
- \omega'^2 \left( 1 - x
      \right) y^2 \left( 1 - y \right)^3 - \omega^2 \left( 1 - x
      \right)^2 y
\left( 1 - y \right)^3
      \left( y + x \left( 1 - y \right) \right)  \right. \right. \nonumber \\
& & \left. + \omega \omega' \left( 1 -
      x \right) y \left( 1 - y \right)^3 \left( y + x \left( 1 - y
      \right) + y \left( 1 - x \right)
\right) 
\right. \nonumber
\\
& & \left. + \frac{1}{2} \left( t - q^2 \right) \left( 1 - x \right)
y \left( 1 - y
  \right)^3  \left( y + x \left( 1 - y \right) + y
    \left( 1 - x \right) \right) \right. \nonumber \\
& & \left. + q^2 \left( 1 - x \right)^2 y \left( 1 - y \right)^3
  \left( y + x \left( 1 - y \right) \right) \vphantom{\frac{1}{2}}
\right) \nonumber \\
& & \left. \times J_0'' \left( \omega \left( y + x \left( 1 - y \right)
  \right) - \omega' y, m_{\pi}^2 - t \left( 1 -
    x \right) y \left( 1 - y \right) - q^2 x \left( 1 - x \right) \left( 1 - y \right)^2
\right) \vphantom{\frac{1}{2}}
\right]
 \nonumber \\
& & + a \cdot q \varepsilon'^* \cdot q \left[
\left( 1 - y \right) \left( y + x \left( 1 - y \right) - \frac{1}{2}
\right) \left( 6 \left( 1 - x \right) \left( 1 - y \right) - \left( y
    + x \left( 1 - y \right) \right) \right)
      \right. \nonumber \\
& & \left. \times J_2'' \left( \omega \left( y + x \left( 1 - y
      \right) \right) - \omega' y, m_{\pi}^2 - t \left( 1 - x \right)
y \left( 1 - y
    \right) - q^2 x \left( 1 -x \right) \left( 1
      - y \right)^2 \right) \right. \nonumber \\
& & \left. + \left( \omega'^2 \left( 1 - x \right)
    y \left( 1 - y \right)^3 \left( y + x \left( 1 - y \right) -
      \frac{1}{2} \right)  \right. \right. \nonumber \\
& & \left. \left. + \omega^2 \left( 1 - x
    \right)^2 \left( 1 - y \right)^3 \left( y + x \left( 1 - y \right) \right)
\left( y + x \left( 1 - y \right) -
      \frac{1}{2} \right) \right. \right. \nonumber \\
& & \left. \left. - \omega \omega' \left( 1 - x \right) \left( 1 - y
    \right)^3 \left( y+ x
\left( 1 - y \right) + y \left( 1 - x
      \right) \right) \left( y + x \left( 1 - y \right) -
      \frac{1}{2} \right)
\right. \right. \nonumber \\
& & \left. \left.
- \frac{1}{2} \left( t - q^2 \right)
\left( 1 - x \right) \left( 1 - y \right)^3
\left( y + x \left( 1 - y \right) + y \left( 1 - x \right) \right) \left( y + x \left( 1 - y \right)
  - \frac{1}{2} \right)
\right. \right. \nonumber \\
& & \left. \left.
- q^2
\left( 1 - x \right)^2 \left( 1
  - y \right)^3 \left( y + x \left( 1 - y
  \right) \right) \left( y + x \left( 1 - y \right) - \frac{1}{2}
\right)
\right) \right.
\nonumber \\
& & \left. \left. \times J_0'' \left( \omega \left( y + x \left( 1 - y
      \right) \right) - \omega' y, m_{\pi}^2 - t \left(
        1 - x \right)
y \left( 1 - y \right) - q^2 x \left( 1 - x \right) \left( 1 - y
      \right)^2 \right) \vphantom{\frac{1}{2}}
\right] \right]
\,
, \\
{\cal{M}}_{VCS}^{(8)} & = & 2 \frac{i e^2 g_A^2}{F^2}
\int_0^1 d x \int_0^1 d
y \nonumber \\
& & \times
\left[ \vphantom{\frac{1}{2}} a \cdot \varepsilon'^*
\left[\vphantom{\frac{1}{2}}
- 5 \left( 1 - y \right) \right. \right. \nonumber \\
& & \left. \times J_6'' \left( \omega y - \omega'
    \left( y + x \left( 1 - y \right) \right), m_{\pi}^2 - t
\left( 1 - x \right) y \left(
      1 - y \right) - q^2 x y \left( 1 - y
    \right) \right) \right. \nonumber \\
& & \left. + \left( \vphantom{\frac{1}{2}}
- \omega^2 y \left( 1 - y \right)^2 - \omega'^2 \left( 1 - x \right)
\left( 1
    - y \right)^2 \left( y + x \left(1-y \right) \right)
\right. \right. \nonumber \\
& & \left. \left.
+ \omega' \omega \left( 1 - y \right)^2 \left( y + x \left(1 -
    y \right) + y \left( 1 - x \right) \right)
+ q^2 y \left( 1 - y \right)^2
\right. \right. \nonumber \\
& & \left. \left. - \frac{1}{2} \left(
      q^2 - t \right) \left( 1 - y \right)^2 \left(y + x \left( 1 - y
      \right) + y \left( 1 - x \right) \right) \right)
\right. \nonumber \\
& & \left. \times J_2'' \left( \omega y - \omega' \left( y + x \left(1-y
\right) \right), m_{\pi}^2 - t \left( 1 - x
\right)
y \left( 1 - y \right)
- q^2 x y \left( 1 - y \right) \right) \vphantom{\frac{1}{2}}
\right] \, \nonumber \\
& & + a \cdot q' \varepsilon'^* \cdot q \left[
  \vphantom{\frac{1}{2}}
\left(
\left(1-x \right) \left( 1 - y \right)^2 \left( 1 - 6 y \right) + y
\left( 1 - y \right) \left( y + x \left( 1 - y \right) \right)
\right)
\right. \nonumber \\
& & \left. \times
J_2'' \left( \omega y - \omega' \left( y + x \left(1-y
\right) \right), m_{\pi}^2 - t \left( 1 - x
\right) y \left( 1 - y \right)
- q^2 x y \left( 1 - y \right) \right) \right. \nonumber \\
& & + \left( \vphantom{\frac{1}{2}}
- \omega^2 \left( 1 - x
  \right) y^2 \left( 1 - y \right)^3 \right. \nonumber \\
& & \left. - \omega'^2 \left( 1 - x \right)^2 y \left( 1 - y \right)^3
  \left( y + x \left( 1 - y \right) \right) \right. \nonumber \\
& & \left. + \omega \omega' \left( 1 - x
  \right) \left( 1 - y \right)^3 y
\left( y + x \left( 1 - y \right) + y \left( 1 - x \right)
  \right) \right. \nonumber \\
& & \left. + q^2 \left( 1 - x \right) y^2 \left( 1 - y \right)^3
\right. \nonumber \\
& & \left. + \frac{1}{2} \left( t - q^2 \right) \left( 1 - x \right) y
  \left( 1 - y \right)^3
  \left( y + x \left( 1 - y \right) + y \left(
      1 - x \right) \right) \right) \nonumber \\
& & \left. \times
J_0'' \left( \omega y - \omega' \left( y + x \left(1-y
\right) \right), m_{\pi}^2 - t \left( 1 - x
\right) y \left( 1 - y \right)
- q^2 x y \left( 1 - y \right) \right) \vphantom{\frac{1}{2}}
\right] \nonumber \\
& & + a \cdot q \varepsilon'^* \cdot q \left[
\left( 1- y \right) \left( 1 - 7 y \right)
\left( y - \frac{1}{2} \right)
\right. \nonumber \\
& & \left. \times J_2'' \left(
\omega y - \omega' \left( y + x \left(1-y
\right) \right), m_{\pi}^2 - t \left( 1 - x
\right) y \left( 1 - y \right)
- q^2 x y \left( 1 - y \right) \right) \right. \nonumber \\
& & + \left( - \omega^2 y^2 \left( 1 - y \right)^2 \left( y -
    \frac{1}{2} \right) \right. \nonumber \\
& & \left. - \omega'^2 \left( 1 - x \right) y \left( 1 - y \right)^2
  \left( y + x \left( 1 - y \right) \right) \left( y - \frac{1}{2}
  \right) \right. \nonumber \\
& & \left. + \omega \omega' y \left( 1 - y \right)^2 \left( y + x
    \left( 1 - y \right) + y \left( 1 - x \right) \right)
\left( y - \frac{1}{2} \right) \right. \nonumber \\
& & \left. + q^2 y^2 \left( 1 - y \right)^2 \left( y - \frac{1}{2}
  \right) \right. \nonumber \\
& & \left. - \frac{1}{2} \left( q^2 - t \right) y \left( 1 - y
  \right)^2
\left( y -
    \frac{1}{2} \right) \left( y + x \left( 1 -
      y \right) + y \left( 1 - x \right) \right) \right) \nonumber \\
& & \left. \left. \times J_0'' \left(
\omega y - \omega' \left( y + x \left(1-y
\right) \right), m_{\pi}^2 - t \left( 1 - x
\right) y \left( 1 - y \right) - q^2 x y \left( 1 - y \right) \right) \vphantom{\frac{1}{2}}
\right] \right] \\
{\cal{M}}_{VCS}^{(9)} & = & \frac{i e^2 g_A^2}{F^2} a
\cdot \varepsilon'^* \int_0^1 d x \left[
\vphantom{\left( \left( \omega' - \omega \right)^2 \left( 1 - x
    \right) x + t x \left( x - 1 \right) \right)}
\left(
    d - 1 \right) J_2' \left( - \left( \omega' - \omega \right) x,
    m_{\pi}^2 - t x \left( 1 - x \right) \right) \right. \nonumber \\
& & + \left. \left( \left( \omega' - \omega \right)^2 x \left( 1 - x
    \right) - t x \left( 1 - x \right) \right) J_0' \left( - \left(
      \omega' - \omega \right) x, m_{\pi}^2 - t x \left( 1 - x \right)
    \right) \right] \label{M_9l} \, .
\end{eqnarray}
\end{mathletters}
Here we have used the definitions
\begin{mathletters}
\begin{eqnarray}
J_0 \left( \omega, m_{\pi}^2 \right) & = & - 4 L \omega +
\frac{\omega}{8 \pi^2} \left( 1 - 2 {\mathrm{ln}} \frac{m_{\pi}}{\mu}
\right) - \frac{1}{4 \pi^2} \sqrt{m_{\pi}^2 - \omega^2}
{\mathrm{arccos}} \frac{-\omega}{m_{\pi}} + {\cal{O}}\left( d - 4
\right) \, , \label{J_0} \\
J_2 \left( \omega, m_{\pi}^2 \right) & = & \frac{1}{d-1} \left[ \left(
    m_{\pi}^2 - \omega^2 \right) J_0 \left( \omega, m_{\pi}^2 \right)
  - \omega \Delta_{\pi} \right]\, , \label{J_2}\\
J_6 \left( \omega, m_{\pi}^2 \right) & = & \frac{1}{d^2 + 6 d + 5} \left( \left( d + 5
  \right) m_{\pi}^2 - 6 \omega^2 \right) J_2 \left( \omega,
  m_{\pi}^2 \right)
\nonumber \\
& &
+ \frac{1}{d^2 + 6 d + 5} \omega^2 \left( \omega^2 -
  m_{\pi}^2 \right) J_0 \left( \omega,
  m_{\pi}^2 \right) \nonumber \\
& & + \frac{1}{d^2 + 6 d + 5} \left(
  \omega^3 - \omega m_{\pi}^2 \left( 1 + \frac{5}{d} \right) \right)
\Delta_{\pi} \label{J_6} \, .
\end{eqnarray}
\end{mathletters}
In Eqs.\ (\ref{J_0}), (\ref{J_2})
and (\ref{J_6}) we have used the same conventions as \cite{BKM1},
\begin{eqnarray}
\Delta_{\pi} & = & 2 m_{\pi}^2 \left( L + \frac{1}{16 \pi^2}
  {\mathrm{ln}} \frac{m_{\pi}}{\mu} \right) + {{O}} \left( d - 4
\right) \, , \nonumber\\
L & = & \frac{\mu^{d-4}}{16 \pi^2} \left[ \frac{1}{d-4} + \frac{1}{2}
  \left( \gamma_E - 1 - {\mathrm{ln}} 4 \pi \right) \right] \, ,
\end{eqnarray}
where we introduced the Euler-Mascharoni constant, $\gamma_E =
0.557215$, and the scale $\mu$ in the dimensional regularization
scheme we use for the evaluation of the integrals.
With $J_{i}'$ and
$J_{i}''$ we denote the first and second partial derivative with respect to
$m_{\pi}^2$,
\begin{mathletters}
\begin{eqnarray}
J_i'\left(\omega,m_{\pi}^2\right) & = & \frac{\partial}{\partial
\left( m_{\pi}^2 \right)} J_i \left( \omega,m_{\pi}^2 \right) \, , \\
J_i''\left(\omega,m_{\pi}^2\right) & = & \frac{\partial^2}{\partial
\left( m_{\pi}^2 \right)^2} J_i \left( \omega,m_{\pi}^2 \right) \, .
\end{eqnarray}
\end{mathletters}
\acknowledgments
The work of GK and SS has been supported by Deutsche
Forschungsgemeinschaft (SFB 201) and Studienstiftung des Deutschen
Volkes.  The research of TRH and BRH is supported in part by the
National Science Foundation.
One of the authors (GK) would like to express his gratitude to
the High Energy Theory and the Nuclear Physics Group at the University
of Massachusetts in Amherst, where the main part of this work was
done, for its kind hospitality.
Moreover, GK and SS are
indebted to Prof. D. Drechsel and A. Metz for valuable
discussions in the final stage of the calculation.

\newpage
\clearpage

\begin{figure}[h]
\centerline{
\epsfxsize=14.5cm
\epsfbox{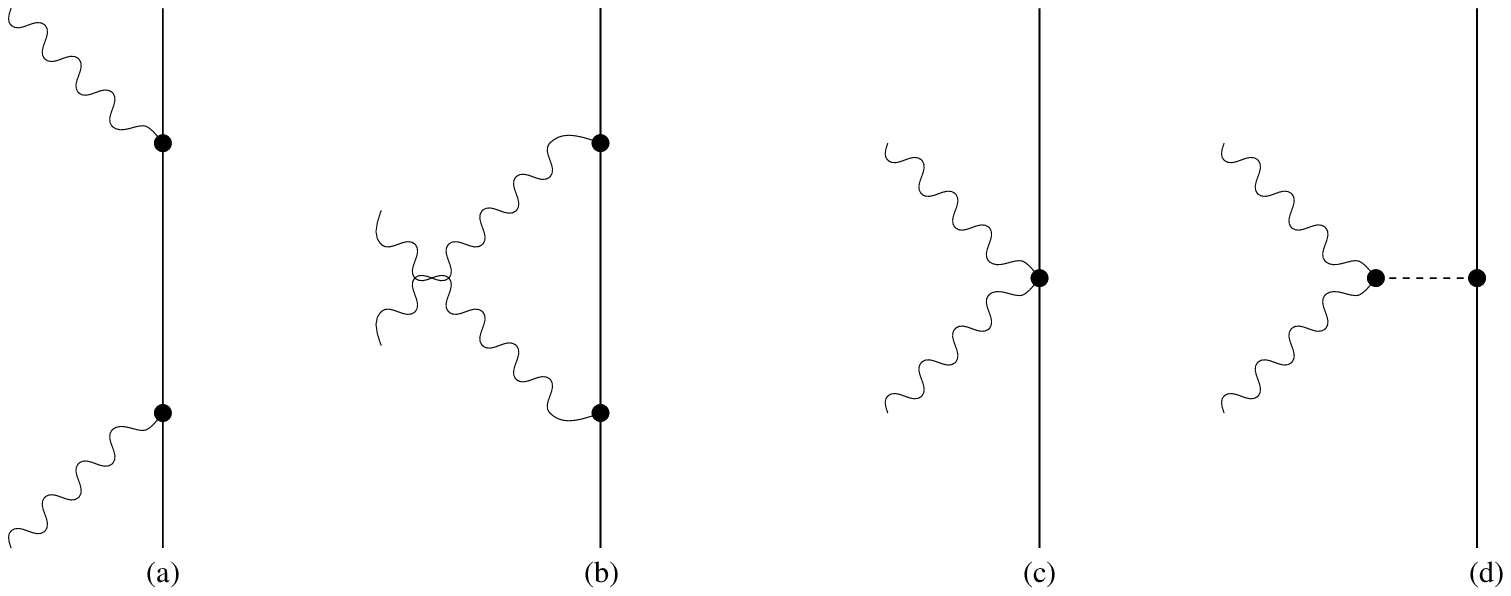}
}
\vspace{0.5cm}
\caption{
\small
Tree diagrams in virtual Compton scattering.}
\label{treefig}
\end{figure}

\newpage
\clearpage

\begin{figure}[h]
\centerline{
\epsfxsize=11.5cm
\epsfbox{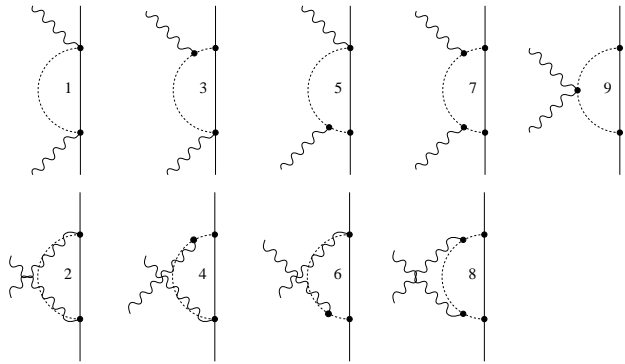}
}
\vspace{0.5cm}
\caption{
\small
Loop diagrams in virtual Compton scattering.}
\label{loopfig}
\end{figure}
\end{document}